\documentclass[twocolumn,showpacs,preprintnumbers,amsmath,amssymb]{revtex4}
\usepackage{graphicx}
\usepackage{subfigure}
\usepackage{dcolumn}
\usepackage{bm}
\usepackage{subeqnarray}

\begin{document}

\title{Nonlocal imaging by conditional averaging of random reference measurements}
\date{20 June 2012}
\author{ LUO Kai-Hong $^1$, HUANG Boqiang $^1$, ZHENG Wei-Mou $^2$, and WU Ling-An $^1$}

\thanks{Corresponding author: wula@aphy.iphy.ac.cn}
\affiliation{$^1$Laboratory of Optical Physics, Institute of Physics
and Beijing National Laboratory for Condensed Matter Physics,
Chinese Academy of Sciences, Beijing 100190, China\\$^2$Institute of
Theoretical Physics, Chinese Academy of Sciences, Beijing 100190,
China}

\begin{abstract}
We report the nonlocal imaging of an object by conditional averaging of the
random exposure frames of a reference detector, which only sees the freely
propagating field from a thermal light source. A bucket detector, synchronized
with the reference detector, records the intensity fluctuations of an identical
beam passing through the object mask. These fluctuations are sorted according
to their values relative to the mean, then the reference data in the
corresponding time-bins for a given  fluctuation range are averaged, to produce
either positive or negative images. Since no correlation calculations are
involved, this correspondence imaging technique challenges our former
interpretations of ``ghost" imaging. Compared with conventional correlation
imaging or compressed sensing schemes, both the number of exposures and
computation time are greatly reduced, while the visibility is much improved. A
simple statistical model is presented to explain the phenomenon.
\end{abstract}


\maketitle

Classical image formation is most commonly realized by recording the
interaction information between a radiation source and the object onto a
detector, all along a single light path. However, in the technique now known as
``ghost" imaging (GI),  two beams are used from the same optical source; one
beam interacts with the target and is collected by a so-called ``bucket"
detector which has no spatial resolution and only measures the total intensity,
while the other propagates through free space to a reference detector which
does have spatial resolution. Neither detector can ``see" the object on its
own, but when their second-order correlation is measured, the image appears in
the coincidence data. This seems quite contrary to intuition, and after the
first experiment was demonstrated with entangled twin beams \cite{Pittman}, the
term ``ghost" was coined to emphasize its peculiar nonlocal nature,  which was
attributed to the quantum properties of the biphoton source. Because of the
potential applications, interest soon spread to other sources \cite{Cheng},
followed by a profusion of studies on GI with pseudothermal light
\cite{Bennink,Gatti,Ferri2005,Cai,Cao,Valencia}, true thermal light
\cite{Zhang}, and even GI without a lens
\cite{Scarcelli,Basano,Meyers,Liu,Ferri,Chen}. The expression ``nonlocal
imaging" is now understood to mean that the light recorded by the reference
detector travels through free space and never interacts with the object, and is
still generally accepted in the GI community irrespective of whether the source
is quantum or classical. Since the prerequisite for correlated imaging is the
identical spatial distribution of the field intensities in the two beams, it
was then realized that the reference beam could just as well be replaced by an
artificially generated random spatial modulation of the object beam, thus only
a bucket detector would be necessary. After the experimental demonstration of
this kind of computational ghost imaging \cite{Shapiro,Bromberg}, attention
turned to improving the techniques involved in the data collection, correlation
and averaging of thousands of exposure frames. Compressed sensing was found to
be very effective in greatly reducing the number of frames required when the
image data is sparse, and has already been implemented \cite{Katz}.

Despite all these advances, there is still much debate on the
various quantum and classical interpretations of GI \cite{
ShihIEEE2007,Shapiro}. Basically, two different interpretations
have been offered: one is multi-photon interference based on
Glauber's quantum optical coherence, and the other is intensity
fluctuation correlation. From the point of view of image
processing, the key to GI reconstruction is the extraction of
information according to the time correspondence between the bucket
signals and reference frames. From the mathematical viewpoint, the
image retrieval is akin to an inverse problem in which the features
(pixels) of an object are to be inferred, given a time series of
bucket and reference light field intensities.

\begin{figure}[tbp]
\includegraphics*[width=0.75\linewidth]{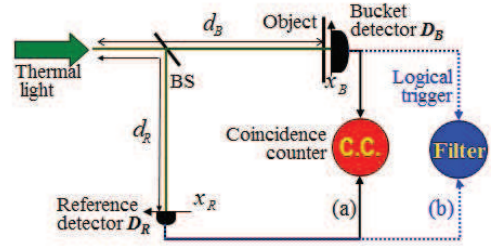}
\caption{Schematic of correlated imaging. BS: beamsplitter. (a)
Conventional GI; (b) Correspondence imaging with selected reference
information.} \label{fig1}
\end{figure}

The basic scheme of a correlated GI system is shown in
Fig.~\ref{fig1}(a). Thermal (or entangled) light is separated into
two parts through a beamsplitter (BS). An object characterized by
the transmission function $T\left( {\vec x_0 } \right)$, where $\vec
x_0$ indicates the transverse coordinate in the plane of the object,
is inserted in front of a bucket detector ${D}_B$ (without any
spatial resolution), which registers the total intensity $I_B$
transmitted (or reflected) by the object, while a reference detector
${D}_R$ with spatial resolution is in the other empty arm. The
distances from the source to the object and to $D_R$ are $d_B$ and
$d_R$, respectively. For lensless GI with thermal light, the
phase-conjugate mirror requirement ${d_B = d_R}$ must be satisfied
\cite{Cao,Scarcelli}. In conventional GI, an image of the mask is
retrieved from the coincidence measurements of ${D}_B$ and ${D}_R$
when the latter is scanned in the corresponding image plane, so all
the information from both the detectors is required. The basis of GI
is the transverse spatial second-order correlation\cite{Glauber}
\begin{eqnarray}
G^{\left( 2 \right)} \left( {\vec x_B ,\vec x_R } \right) & =
&\frac{1}{N}\sum\limits_{\{t_i \mid i = 1, \cdots, N\}}^{} {I_B
\left( {\vec x_B ,t_i } \right)I_R \left( {\vec x_R ,t_i } \right)}.
\label{eq:secondcorrelation1}
\end{eqnarray}
Here $ I_B \left( {\vec x_B }, t_i \right)$, $ I_R \left( {\vec x_R
}, t_i \right)$  and $\vec x_B$, $\vec x_R$ are the intensities at
time $t_i$ and the transverse coordinates at ${ D}_B$ and ${ D}_R$,
respectively; $N$ is the number of the coincidence measurements,
i.e. exposure frames. The output ${ I}_B(t_i)$ of the bucket
detector is the spatial integral $\int {d\vec x_B}I_B \left( {\vec
x_B }, t_i \right)$, which erases the spatial resolution of the
target.

Below, we demonstrate a method using incoherent thermal light by
which an image of the object can be obtained very simply from the raw
reference detector data, without the need of any complicated image reconstruction method
such as that based on compressed sensing \cite{Wu1}. In this method, which we
shall call correspondence imaging (CI), we only have to perform
signal averaging over those reference frames selected by an
appropriate gate from the bucket detector. Since no direct second-order
correlation is involved, the theory seems to challenge all our
previous interpretations of GI. In addition to the normal positive
image, a reversed negative image can also be obtained, and the
quality of the images may be far superior to that of conventional
GI for the same number of frames.

A schematic of the CI setup is shown in Fig.~\ref{fig1}(b). As in (a), a linearly polarized He-Ne laser beam diameter 5 mm is projected onto a
rotating ground-glass disk to produce pseudothermal light
\cite{Goodman}, which is then split by a 50:50
non-polarizing BS into two beams, one of which passes through an
object mask to detector $D_B$, and the other goes directly to $D_R$,
to produce the bucket intensity $I_B$ and the reference intensity
distribution $I_R \left( {\vec x_R } \right)$, respectively. The
distance from the ground-glass disk to BS is 19 cm, and from BS to
the mask 11 cm, that is, $d_B=d_R \sim 30$ cm. Both detectors
(Imaging Source DMK 31BU03) are synchronized by the same trigger
pulse at a rate of about 3 Hz. The data was acquired with an
exposure time ($0.1$ ms) much shorter than the coherence time of the
laser, and saved to a computer. A total of $50,000$ frames
was grabbed by each camera per image plot. The bucket detector could
also be just a point detector so long as it collects all the light
transmitted through the mask, for example by means of a focusing
lens.

\begin{figure}[tbp]
\includegraphics*[width=0.8\linewidth]{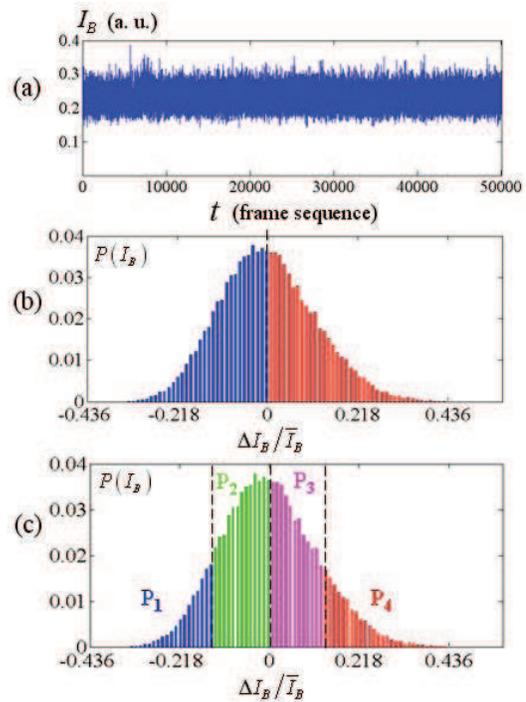}
\caption{Bucket detector intensity ${I_B}$. (a) Integrated beam
intensity ${I_B}$ recorded as a function of time sequence. (b)
Probability distribution of $I_B$ centered about its mean. The left
(blue) block is binned for $\Delta I_B < 0$, and the right (red)
block for $\Delta I_B  \ge 0$. The dashed line denotes $\bar I_B$.
(c) Multi-section binning. From left to right, the dashed lines
denote $I_N$, $\bar I_B$ and $I_P$, where $I_N$ and $I_P$ satisfy
$P(I_N) = P(I_P ) = P(\bar I_B)/2$.} \label{fig2}
\end{figure}

The integrated beam intensity ${I_B(t_i)}$ at the bucket detector
recorded as a function of time is shown in Fig.~\ref{fig2}(a); it
can be seen that the intensity fluctuations are random. We then
calculate the deviation from the mean $\Delta I_B(t_i)  = I_B(t_i) -
\bar I_B$, and plot the histogram of $I_B(t_i)$, or equivalently of
the normalized intensity fluctuations $\Delta I_B(t_i)/\bar I_B$, to
show the probability distribution $P(I_B)$ of $I_B$. Using the
intensity $I_B(t_i)$ at time $t_i$ as an indicator, we divide
$\{t_i\}$ into two subsets:
\begin{eqnarray}
\label{eq:timeseparation}
t_- =
\left\{ {t_i \left| {I_B \left( {t_i } \right) < \bar I_B}
\right.} \right\}, \
t_+ = \left\{ {t_i \left| {I_B \left(
{t_i} \right) \ge \bar I_B} \right.} \right\}.
\end{eqnarray}
All the ${I_B(t_i)}$ with ${t_i} \in {t_-}$ (or ${t_+}$) contribute
to the left (or right) half of Fig.~\ref{fig2}(b).

Because both detectors are synchronized in time, the reference
signals $I_R \left( {\vec x_R}, t_i \right)$ may also be
divided into two groups according to $t_i \in t_+$ or $t_i \in t_-$.
This is equivalent to labeling $I_R \left( {\vec x_R}, t_i \right)$
with the time stamp of ${I_B(t_i)}$. By dividing the range of $I_B$
into more sections, say at $I_N$ and $I_P$ as well as at $\bar I_B$,
with $\Delta I_N < 0$, $\Delta I_P > 0$, and $P(I_N) = P(I_P ) =
P(\bar I_B)/2$, we can perform multi-section binning for $I_B$
(Fig.~\ref{fig2}(c)), and hence, correspondingly, for $I_R
\left({\vec x_R}, t_i \right)$.

\begin{figure}[tbp]
\includegraphics[scale=0.4]{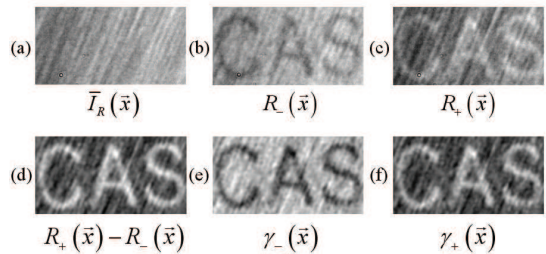}
\caption{Images obtained by nonlocal CI; all figures are
automatically gray-scale compensated. (a) Average of all the frames
from $D_R$. (b) Negative and (c) positive correspondence images
after time binning of $D_R$ based on negative and positive $\Delta
I_B$, respectively. (d) Images obtained using all the frames from
$D_R$ but with the negative image signals ${R_{ -} \left( {\vec x }
\right)}$ inverted. (e) and (f), as in (b) and (c) above,
respectively, but after normalized array division by (a). }
\label{fig3}
\end{figure}

Due to the free propagation of the thermal light field, each
reference frame is purely random, so the average value of all the
reference signals is
\begin{eqnarray}
\bar I_R \left( {\vec x_R } \right)  = \frac{1}{N}\sum\limits_{\{t_i
\mid i = 1, \cdots, N\}}^{} {I_R \left( {\vec x_R ,t_i } \right)}=
\bar I_R. \label{eq:wreferenceaverage1}
\end{eqnarray}
Statistically, the mean value of a random variable is a constant, so
we would not expect $\bar I_R \left( {\vec x_R } \right)$ to reveal
anything, however long the exposure time, as shown in
Fig.~\ref{fig3}(a), and for a homogeneous thermal field each pixel
would have the same averaged intensity. But amazingly, after the
above partitioning, the target in the object arm can be
reconstructed merely by taking the average of only the reference
intensities within a selected distribution:
\begin{eqnarray}
R_{\pm}
\left( {\vec x_R } \right) & = & \frac{1}{N_{\pm}}\sum\limits_{\{t_i
\mid t_i \in t_{\pm}\}}^{} {I_{R } \left( {\vec x_R
,t_i } \right)},
\label{eq:referenceimaging}
\end{eqnarray}
where $N_ \pm = \sum _ {t_i \in t_\pm} 1$ is the cardinality of $t_
\pm$ (see Figs.~\ref{fig3}(b) and (c)).

We know from information theory and statistics that a complete
random basis is recommended for image reconstruction. In our
experiment this is supplied by the spatial fluctuations of all the random modes of the
thermal beams. The
object and $D_R$ are at the same distance from
the source, so they see the same mode distribution. Part of this
field, i.e. the transmitted (or reflected) light through the mask,
contributes to the bucket signal, and may fluctuate above or below
its mean value in each exposure. The intensity of the corresponding
$D_R$ pixels will also fluctuate in harmony, but the
rest of the frame will be fluctuating in an uncorrelated way. Let us
divide the spatial pixels at the mask $\{ \vec x_0 \}$ into subsets
$X_1$ and $X_0$ with the former transmitting and the latter blocking
light. It is the pixels $\{\vec x_0 \mid \vec x_0 \in {X_1}\}$ that
contribute to the bucket signal $I_B$ , and when the field intensity
fluctuates higher (lower) in this area, the bucket output will
naturally increase (decrease). The intensity of the corresponding
reference detector pixels will also increase (decrease), but as this
contribution is superimposed upon the entire beam intensity, i.e.
all the light from both areas $X_1$ and $X_0$, the total reference
output will be more or less constant. By separately collecting and
averaging over a sufficiently large number of exposures according to
the positive (negative) bucket signals, the positive (negative)
image of the object ($X_1$) will then stand out from the constant
background.

A conceptual mathematical model is as follows. Since the field distribution
at the mask is the same as that at $D_R$, the bucket intensity may
be written
\begin{equation}
I_B \left( {t_i } \right) = \hbox{$\sum_{\{\vec x \}}$} {I_R \left(
{\vec x ,t_i } \right)\left| {T\left( {\vec x } \right)} \right|^2
}, \label{eq:bucketi1}
\end{equation}
where for convenience we have dropped the subscripts of $x$. From
Eq.~(\ref{eq:wreferenceaverage1}) we thus have
$\bar I_B = \bar I_R \sum\limits_{\{\vec x \}} {\left| {T\left( {\vec
x } \right)} \right|^2 }$.
For a narrow bucket intensity bin, centered at some ${I_B(t_i)}
\approx I^+_B$ far above $\bar I_B$, we may make the following
approximation
\begin{eqnarray}
I_B \left( {t_i } \right) & \simeq & \sum\limits_{\{\vec x \mid \vec
x \in {X_1}\}} {I_R \left( {\vec x ,t_i } \right)\left| {T\left(
{\vec x } \right)} \right|^2 }. \label{eq:bucketi2}
\end{eqnarray}
Since $N_{B+}$ is the number of ${I_B(t_i)}$ values registered in the range of
$I^+_B$, we can write
\begin{eqnarray}
G_+(\vec x )&\equiv&\frac{1}{{N_{B+ } }}\sum\limits_{\{t_i \mid I_B \left( {t_i }
\right) \approx I_B^ +  \}} {I_B
\left( {t_i } \right)I_R \left( {\vec x ,t_i } \right)} \nonumber \\
& \simeq & \frac{{I_B^ +  }}{{N_{B+ } }}\sum\limits_{\{t_i \mid I_B \left( {t_i
} \right) \approx I_B^ + \}} {I_R \left( {\vec x ,t_i } \right)}
\equiv \frac{I_B^ +}{N_B^ +}I^+(\vec x ).
\label{eq:bucketi3}
\end{eqnarray}
On the other hand, by using Eq.~(\ref{eq:bucketi1}), we
have
\begin{eqnarray}
G_+(\vec x ) \simeq   \bar I_B \bar I_R + \Delta _R^2 \left| {T\left( {\vec x
} \right)} \right|^2, \label{eq:bucketi4}
\end{eqnarray}
where $ \Delta _R^2 = \frac{1}{N_ {B+} } {\sum\limits_{\{t_i \mid I_B \left(
{t_i } \right) \approx I_B^ +\}}} {[I_R(\vec x, t_i)  - \bar I_R]^2 } $. This
means that, for ${\vec x  \in X_1}$
\begin{eqnarray}
I^+(\vec x ) \propto C_b + \left| {T\left( {\vec x } \right)} \right|^2,
\label{eq:bucketip1}
\end{eqnarray}
with $C_b = \bar I_B \bar I_R / \Delta _R^2 $ being a constant due
to the background. Similarly, for ${\vec x} \in {X_0}$ we have
$I^+(\vec x )\propto C_b + 0$. For a narrow bin centered at some $I^-_B$ far below $I_B$, the same
argument is applicable under the exchange $X_1 \leftrightarrow X_0$.
Thus, while the summation of reference frames corresponding to $
{I_B(t_i)} \approx I^+_B$ gives the positive image of pixels in
$X_1$, the summation frames for ${I_B(t_i)} \approx I^-_B$ gives the
negative image of pixels in $X_0$.

When going from the narrow bin at $I^+_B$ to a wide bin of $I_B
\left( {t_i } \right) \ge \bar I_B$, errors from two sources come
in: since $I_B \left( {t_i } \right) \ge \bar I_B$ is approximated
by the single value $I^+_B$, the validity of Eq.~(\ref{eq:bucketi2})
becomes weaker. This implies that the reference frames with their
corresponding $I_B \left( {t_i } \right)$ close to $\bar I_B$
contain limited information about $\left| {T\left( {x } \right)}
\right|^2$, relative to the background, so the image quality is
reduced. However, if we take $R_+(\vec x)-R_-(\vec x)$ then the
background can be removed. We may also define the normalized
distributions $\gamma_ {\mp} ({\vec x }) = R_{ \mp } ({\vec x }) /{\bar I_R
({\vec x }) }$.

Figure \ref{fig3} shows the images obtained for the letters ``CAS" in an object
mask which is 1.4 $\times$ 2.56 mm$^2$ in size (300$\times$ 550 pixels). We
define two levels of transmission, $1$ for total transmission through the
letters, and $0$ elsewhere. The results obtained by various different methods
are presented for comparison to illustrate the advantages of CI. We see from
Fig.~\ref{fig3}(a) that, as expected, merely taking the time average $\bar I_R
\left( {\vec x } \right)$ using all the $50,000$ frames of the reference
detector produces nothing (the slanted lines are artifacts from the rotating
ground glass plate due to the long time exposures). After partitioning by the
two complementary time series given by
Eq.~(\ref{eq:timeseparation}), there
emerges a negative image ${R_{ - } \left({\vec x } \right)}$ from $26,005$
reference frames (Fig.~\ref{fig3}(b)), and a positive image ${R_{ +} \left(
{\vec x } \right)}$ from the remaining frames (Fig.~\ref{fig3}(c)). The image
${R_{ +} \left( {\vec x } \right)} - {R_{ -} \left( {\vec x } \right)}$
obtained using all the information from $D_R$ but with the negative image
signals inverted is shown in Fig.~\ref{fig3}(d). When we take the normalized
correspondence images from $D_R$, obtained as $\gamma_- ({\vec x })$ and
$\gamma_+ ({\vec x })$ by using matrix array division, we obtain the images in
(e) and (f), respectively, which are much clearer than in (b) and (c) above. It
is interesting to note that in all the upper row figures, there is a small
black dot with a white centre at the bottom of the letter ``C", due to
diffraction from a dust particle on the surface of the reference detector.
However, this is no longer visible in the lower row as the background has been
removed through normalization.

It should be noted that although CI still requires synchronization
with the bucket detector signals, it is quite different from
the usual correlated imaging of GI. Each individual frame of $D_R$
is random and the object cannot be seen from the
mean of the total, but its image can be retrieved from the
conditional average of subsets of the data, a seeming
contradiction. Moreover, the visibility of the
images may be optimized by appropriate selection and
weighting of the data partitioning, for example, through appropriate choice of
bucket detector intensities $I_N$ and $I_P$, we can divide the histogram into
four areas, as illustrated in Fig.~\ref{fig2}(c). Besides $\bar I_B$, we
introduce two more $I_N~(< \bar I_B)$ and $I_P~(> \bar I_B)$ to divide the
range of $I_B$ into four subintervals: $P_1 = [0, I_N) $, $P_2 = [I_N, \bar
I_B)$, $P_3 = [\bar I_B, I_P )$, and $P_4 = [I_P, +\infty)$. Assuming that there
are $N_i$ measurements (or frames) with a bucket intensity within the range
$P_i$ ($i = 1, 2, 3,4$), in the same way as we defined ${R_{\mp} \left( {\vec x
} \right)}$, we introduce
\begin{eqnarray}
R_{i} ({\vec x })  & = & \frac{1}{{N_ i }}\sum\limits_{ \{t_i \mid
I_B(t_i) \in P_i \} } {I_{R } \left( {\vec x ,t_i } \right)},
\label{eq:mri1}
\end{eqnarray}
and its normalized form  $\gamma_{i} ({\vec x })=R_{i} ({\vec x }) /{\bar
I_R ({\vec x }) }$. We expect that a negative image would be obtained with $R_1$ or
$R_2$ and a positive image with $R_3$ or $R_4$.

\begin{figure}[tbp]
\includegraphics[scale=0.42]{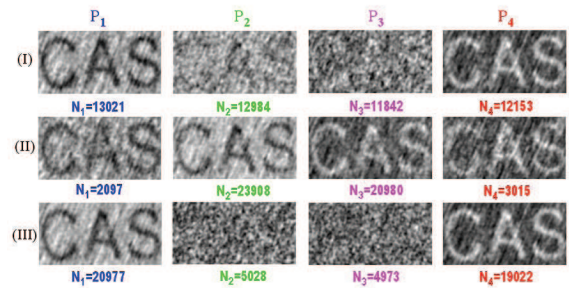}
\caption{Selected CI images when the information from $D_R$ is
divided into four subsets with different weightings (the gray-scale
has been automatically adjusted). (I) Each subset is composed of the
same number of measurements, $N_1 \sim N_2 \sim N_3 \sim N_4$; (II)
most of the frames are concentrated in the central areas $P_2$ and
$P_3$, with a weighting of $N_1:N_2:N_3:N_4 \sim 1:9:9:1$; (III)
when $N_1:N_2:N_3:N_4 \sim 4:1:1:4$, no images can be deciphered
from $P_2$ and $P_3$.} \label{fig4}
\end{figure}

Using the same experimental data as above, three sets of values for
$N_i$ were chosen to create the correspondence images of
Fig.~\ref{fig4}. In the first row, the number of reference frames
contained in each sector $P_i$ was the same ($N_1 \sim N_2 \sim N_3
\sim N_4$). It is obvious that the images obtained from $P_1$ and
$P_4$ are far superior in quality compared to those from $P_2$ and
$P_3$, although the same number of frames were used. In the second
row, a weighting of $N_1:N_2:N_3:N_4 \sim 1:9:9:1$ was used. Even in
this case, we see that the reconstructions from the extrema, i.e.
the wings of the fluctuation distributions in Figs.~\ref{fig2}(b)
and (c), are comparable with those from the central area, despite
the fact that only 5\% of the data was used. In the third row, the
ratios of the four parts are about $N_1:N_2:N_3:N_4 \sim 4:1:1:4$,
and we can no longer see any image in the regions of $P_2$ and
$P_3$. This indicates that most of the object information is
concentrated at the two ends $P_1$ and $P_4$, i.e., the reference
frames corresponding to larger bucket intensity fluctuations contain
more information and contribute more to image retrieval. When the
image is indistinguishable from the background, we may infer that
the exposure intensities resulted from both subsets $X_1$ and $X_0$.
By refining our weighting of the partition selection, the image may
be recovered faster and with fewer measurements, in agreement with
the experimental observations of Fig.~\ref{fig4}.

It should be noted that in Figs.~\ref{fig3} and \ref{fig4} the
gray-scale has been automatically adjusted by the software for better
pictorial detail: the minimum of the image matrix is displayed
as the darkest and the maximum as the brightest so that contrast is
artificially improved. In this way, images that originally had very
different visibilities may look quite similar on a computer
screen, or vice versa. The CI images appear comparable
to those obtained by conventional GI although only half the data
from $D_R$ or even less is used.

\begin{figure}[tbp]
\includegraphics[scale=0.42]{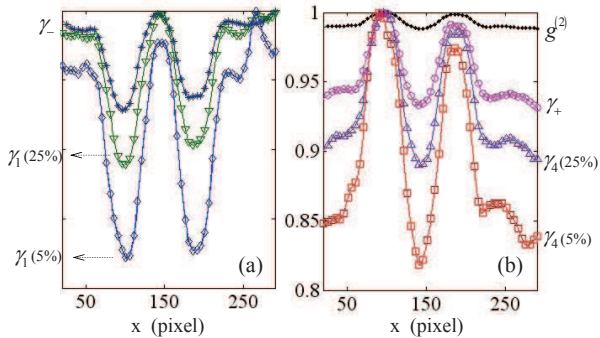}
\caption{Image intensity profiles for a double-slit object;
transverse coordinates are the pixels of $D_R$ (each pixel is 4.65
by 4.65 $\mu$m$^2$). Each plot is normalized by its maximum value.
(a) Normalized negative correspondence images reconstructed for
$\gamma _-$ and two $\gamma _1$ values. (b) Top curve: second-order
normalized GI; lower three curves: normalized correspondence images
reconstructed for $\gamma _+$ and two $\gamma _4$ values.}
\label{fig5}
\end{figure}

To demonstrate the superior visibility of CI more clearly we also performed an
experiment under the same conditions as before but with a simple double-slit as
object (slit width and distance 0.2 mm and 0.4 mm, respectively). The
fluctuation histogram of $I_B$ was divided into four sectors, and to spotlight
the differences in visibility we look at the intensity profiles along one
transverse dimension. For a better comparison of the original image contrast of
the various techniques, in Fig.~\ref{fig5} we have adopted a common gray-scale
in which the image matrices are divided by their maximum value and displayed
with zero as the darkest and 1 as the brightest. Fig.~\ref{fig5}(a) shows the
normalized negative correspondence images, where $\gamma _-$ indicates the
negative image when the probability histogram is only divided into two parts,
and the percentage values in $\gamma _1 (25\%)$ and $\gamma _1(5\%)$ represent
the percentage $N_1$ of reference frames selected in partition $P_1$.
Fig.~\ref{fig5}(b) shows positive images: the top curve is from normalized
second-order  correlation GI; $\gamma _+$ is the normalized positive CI image,
and $\gamma _4(25\%)$ and $\gamma _4(5\%)$ are the CI plots for $N_4/N \simeq
25\%$ and 5\%. The superiority of the latter is immediately obvious. Moreover,
the less the number of exposures that are taken, the better is the visibility
relative to that in conventional GI. The definition and analysis of the
visibility and signal-to-noise ratio in CI is a complicated problem
\cite{Basano1,Brida}, so a more quantitative discussion, as well as a full
theory of CI in general, will be presented in a future paper.

In summary, we have demonstrated
the reconstruction of positive and negative images of a nonlocal
object through selective averaging of the exposures of a reference
detector that has never interacted with the target field.
A simple statistical model is proposed to
explain the phenomenon, which, to our knowledge, cannot be explained
by conventional classical or quantum wave optics. Since the method
is based on a form of conditional averaging of the reference data,
much less information is required than in GI, which in fact contains
much redundant information. Moreover, the visibility is
significantly better, especially if the number of exposure frames is
limited. The reconstruction process only involves straightforward
selection and stacking, so it is much simpler than conventional GI
or compressed sensing methods; complicated calculations and
algorithms are not required, and storage space, memory consumption
and processing time are greatly reduced, which is a particular
advantage when the images are large. The basic concept may also be
extended to similar experiments that use correlation calculation,
including computational GI, so potential applications are plentiful.
Although this CI technique is apparently simple, the underlying
physics is quite subtle and deserves further exploration. It could
open up new opportunities in the field of metrology,
positioning, and imaging.

This work was supported by the National Natural Science Foundation
of China Grant No. 60978002, the National Program for Basic
Research in China Grant Nos. 2010CB922904 and 2007CB814800,
and the National High Technology R \& D Program of China Grant No. 2011AA120102.

\end{document}